\documentclass[
    ,final            
  ]
  {aipproc}

\layoutstyle{8x11single}

\begin{document}

\title{NEW RESULTS FROM THE B-FACTORIES BELLE and BABAR}

\classification{<Replace this text with PACS numbers; choose from this list:
                \texttt{http://www.aip..org/pacs/index.html}>}

\author{G.Leder for the BELLE collaboration}{
  address={Institute for High Energy Physics, Austrian Academy of Sciences, Nikolsdorfergasse 18,A-1050 Vienna,AUSTRIA}
}



\begin{abstract}
 The BELLE detector has already accumulated $e^+ e^-$ collision data at the KEKB-collider corresponding to 600 $fb^{-1}$. BaBar has accumulated data of approximately 390 $fb^{-1}$ at PEP-II. Both are running on asymmetric energy $e^+ e^-$ colliders at the  $\Upsilon(4S)$ energy. The paper selects important results from both experiments with the emphasis on CP violation in B meson decays and its implications for the Unitarity Triangle. 
\end{abstract}

\maketitle


\section{1. Introduction}

 An irreducible complex phase in the weak interaction quark-mixing(CKM) matrix of the Standard Model(SM) describes CP violation(\cite{CKM:1}). The constraint between first and third generation is given by the equation $V_{ud}V_{ub}^*+V_{cd}V_{cb}^*+V_{td}V_{tb}^* = 0 $. This equation can be visualized in the form of the "Unitary triangle" in the complex plane and we label the three angles of this unitary triangle as $\phi_1,\phi_2$ and $\phi_3$. \footnote{$\phi_1=\beta,\phi_2=\alpha$  and $\phi_3=\gamma$ in a different convention} B-factories using $\Upsilon(4S) \rightarrow B^0 - \bar B^0 $ production measure CP violation arising from the interference between $B^0 - \bar B^0$ mixing and decay.    

\section{2. $\phi_1 (\beta)$ measurement }

\subsection{2.1 Time dependent asymmetry}

If both $B^0$ and $\bar B^0$ decay to a common CP eigenstate $"f_{CP}"$, we can define the  time dependent CP asymmtry $A_{CP}(t)$ as
\begin{equation}
A_{CP}(t)=\frac{\Gamma(\bar B^0 \rightarrow f_{CP})-\Gamma(B^0 \rightarrow f_{CP})}{\Gamma(\bar B^0 \rightarrow f_{CP})+\Gamma(B^0 \rightarrow f_{CP})}
         = S_{f_{CP}}sin\Delta mt+ A_{f_{CP}}cos\Delta mt
\end{equation} 
where $\Gamma(\bar B^0|(B^0) \rightarrow f_{CP})$ is the decay rate for a $\bar B^0|(B^0)$ to decay to $f_{CP}$ at a proper time t after $\Upsilon(4S)$-production while  $\Delta m$ is the mass difference between the two $B^0$ mass eigenstates.

\subsection{2.2 $b \rightarrow c \bar c s$ decay} 

\begin{figure}
    \includegraphics[width=6.0cm]{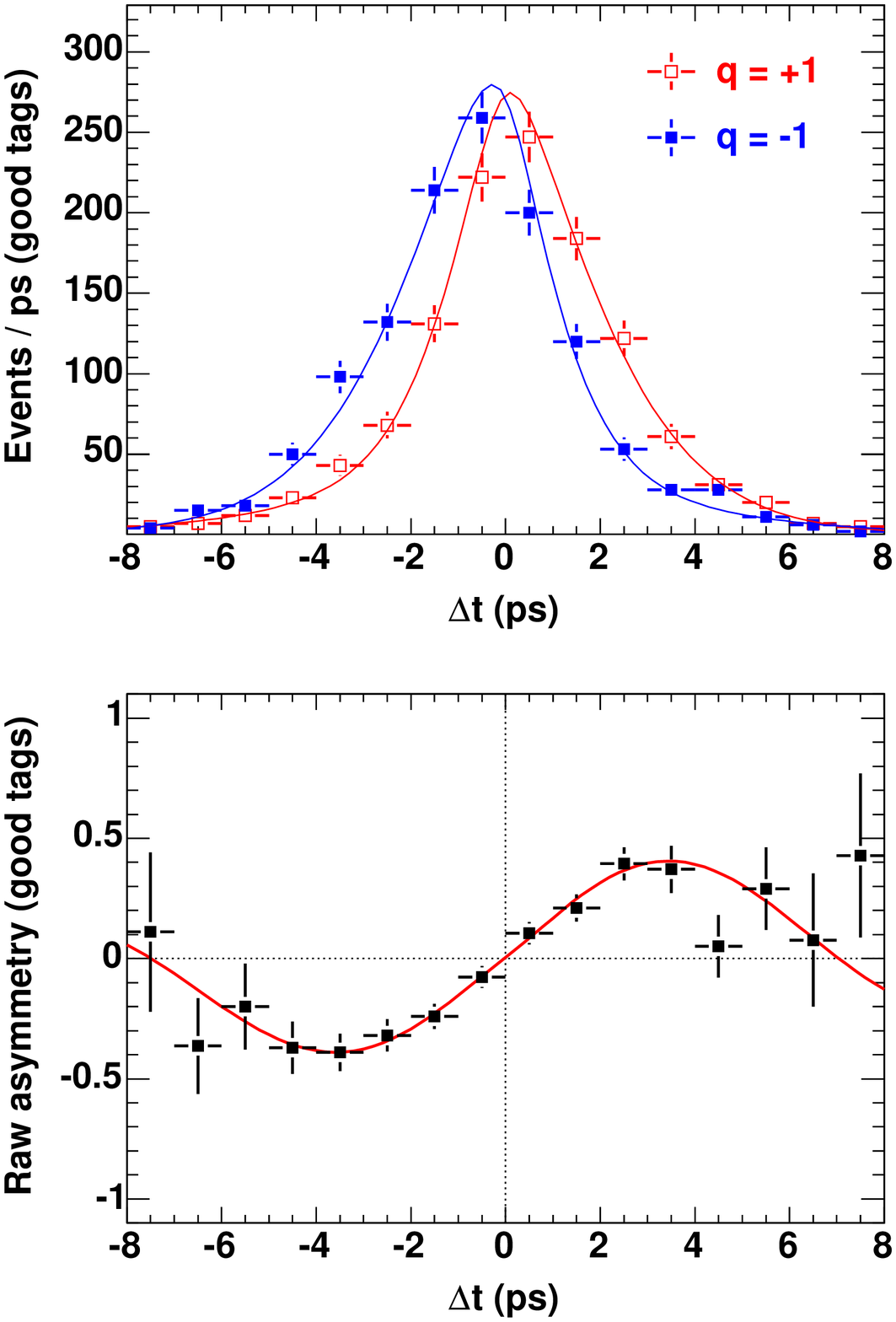}\includegraphics[width=6.0cm]{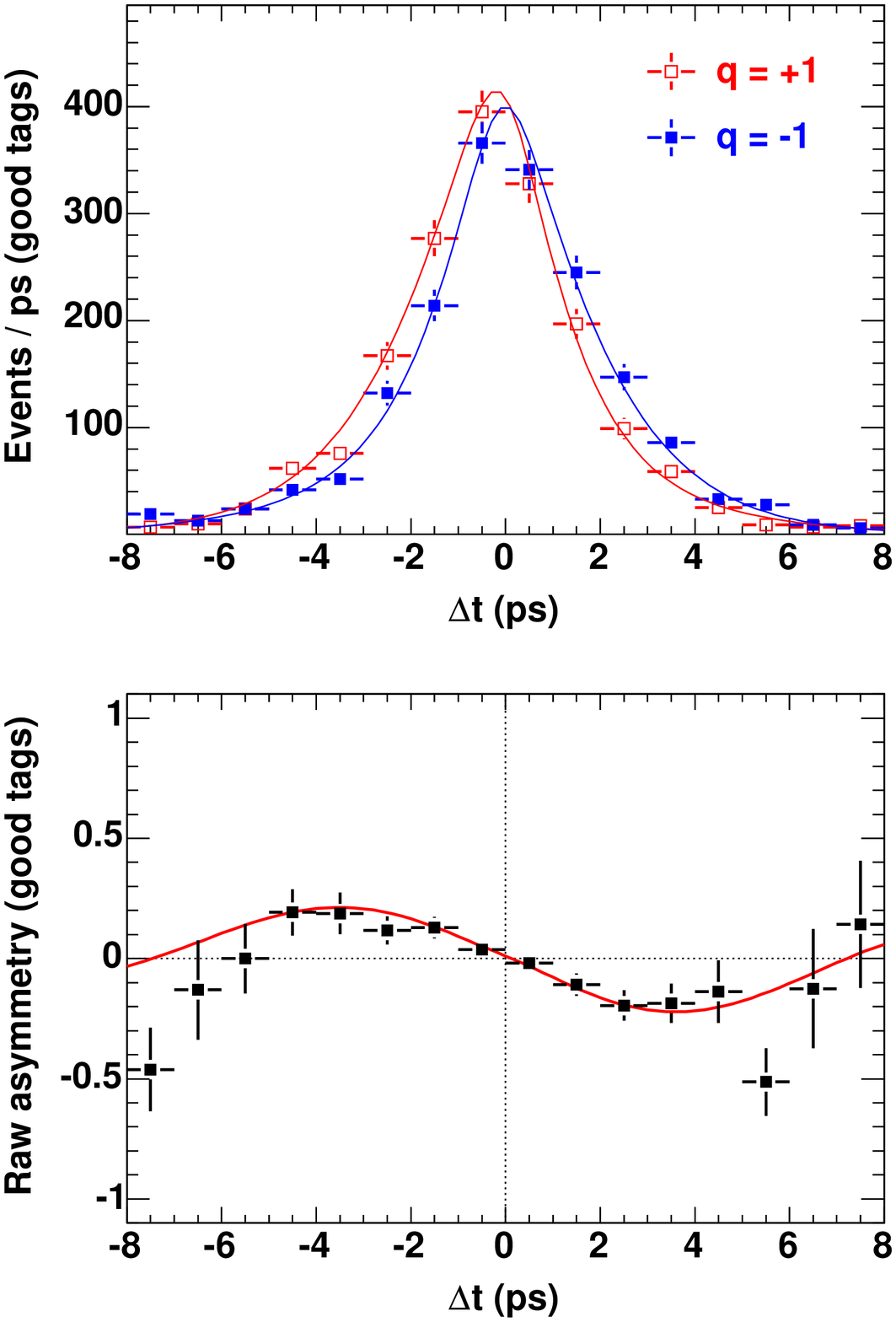}\includegraphics[width=6.0cm]{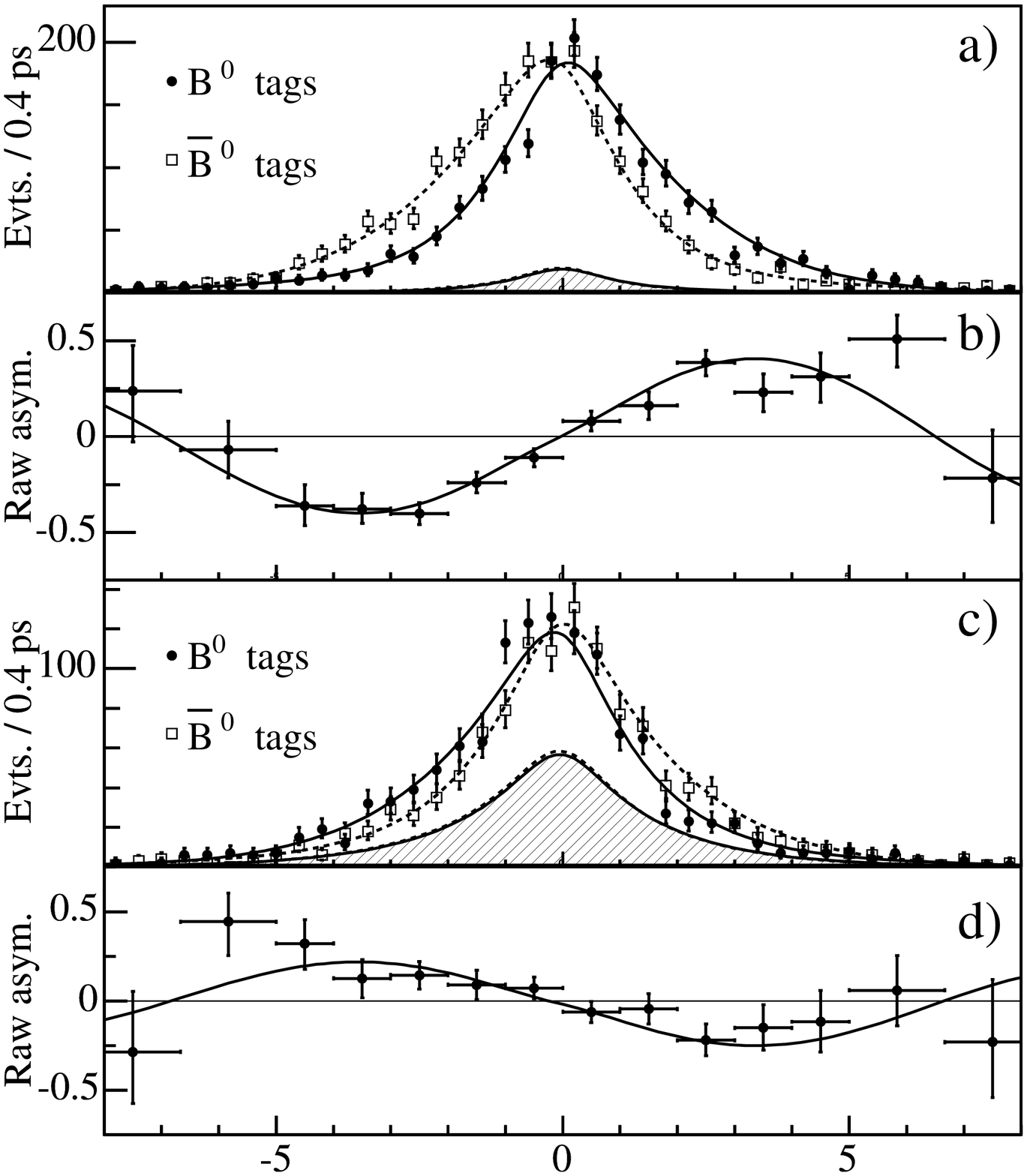}
  \caption{Measurement of $S_{J/\psi K^0}$ from BELLE (left, center) and BABAR (right). Upper part: proper time distribution $\Delta t$ and lower part: raw asymmetry for (left) $B \rightarrow J/\psi K_S$ and (right) $B \rightarrow J/\psi K_L$ for BELLE. For BABAR the two upper plots show $B \rightarrow J/\psi K_S$, while the two lower plots show $B \rightarrow J/\psi K_L$ }
 \label{fig:1}
\end{figure}

This quark transition is the best place to measure $\phi_1$ because of its small theoretical uncertainty. The CP-violation parameters are $S_{b \rightarrow c \bar c s} = - \xi sin2\phi_1$ with $\xi$ the CP-eigenvalue of the final states, and $A_{b \rightarrow c \bar c s} = 0$. The "golden channel" $B \rightarrow J/\psi K^0$ gives the best results for $\phi_1$ because of the good signal-to-noise ratio. From 386M $B\bar B$ events BELLE \cite{BELLE-beta:2} has selected 5,264 $B \rightarrow J/\psi K_S$ and 4792 $B \rightarrow J/\psi K_L$ candidates. The updated result \\
$S_{J/\psi K^0} = 0.652 \pm 0.039(stat) \pm 0.020(syst)$\\
$A_{J/\psi K^0} = 0.010 \pm 0.026(stat) \pm 0.036(syst)$\\  
BABAR  \cite{BABAR-beta:3} has started with 227M $B\bar B$ events and selected 7730 events in both modes. This updated result is now $sin2\beta = 0.722 \pm 0.040(stat) \pm 0.023(syst)$.
The direct CP-violation is $A_{J/\psi K^0}$ is consistent with 0. Fig. ~\ref{fig:1} shows proper time distributions and raw asymmetries for both $B \rightarrow J/\psi K_S$
and $B \rightarrow J/\psi K_L$ measured in both experiments. The different sign of raw asymmetries stems from the different CP-eigenvalues of $K_S$ and $K_L$.

\subsection{2.3 Resolving the $\phi_1$ ambiguity}
Although the CP violating parameter $\phi_1$ can be measured with relatively high precision there remains still a four-fold ambiguity for $\phi_1 $. The Standard Model predicts a positive value of $cos2\phi_1$, and BELLE \cite{BELLE-cos:4} has obtained the best result with a time dependent Dalitz analysis of the neutral D meson in the channel $B \rightarrow D^{(*)}_{CP} \pi^0 (\eta,\omega)$, with the $D_{CP}$ decaying to $K_S\pi^+\pi^-$ Using 386M $B\bar B$ events BELLE obtained \\ 
$sin2\phi_1 = 0.78 \pm0.44(stat)\pm0.22(syst)$  and 
$cos2\phi_1 =+1.87 ^{+0.40}_{-0.53}(stat) ^{+0.22}_{-0.32}(syst)$\\
The negative $cos2\phi_1$ solution is disfavoured with ~2$\sigma$ significance.
BABAR \cite{BABAR-cos:5} has used a different channel and gets $cos2\beta = +2.72 ^{+0.50}_{-0.79}(stat)\pm0.27(syst)$ thus supporting this solution of $\phi_1$ ($\beta$)

\subsection{2.4 $b \rightarrow q \bar q s $ decay modes and the influence of penguins}

\begin{figure}
    \includegraphics[width=6.0cm]{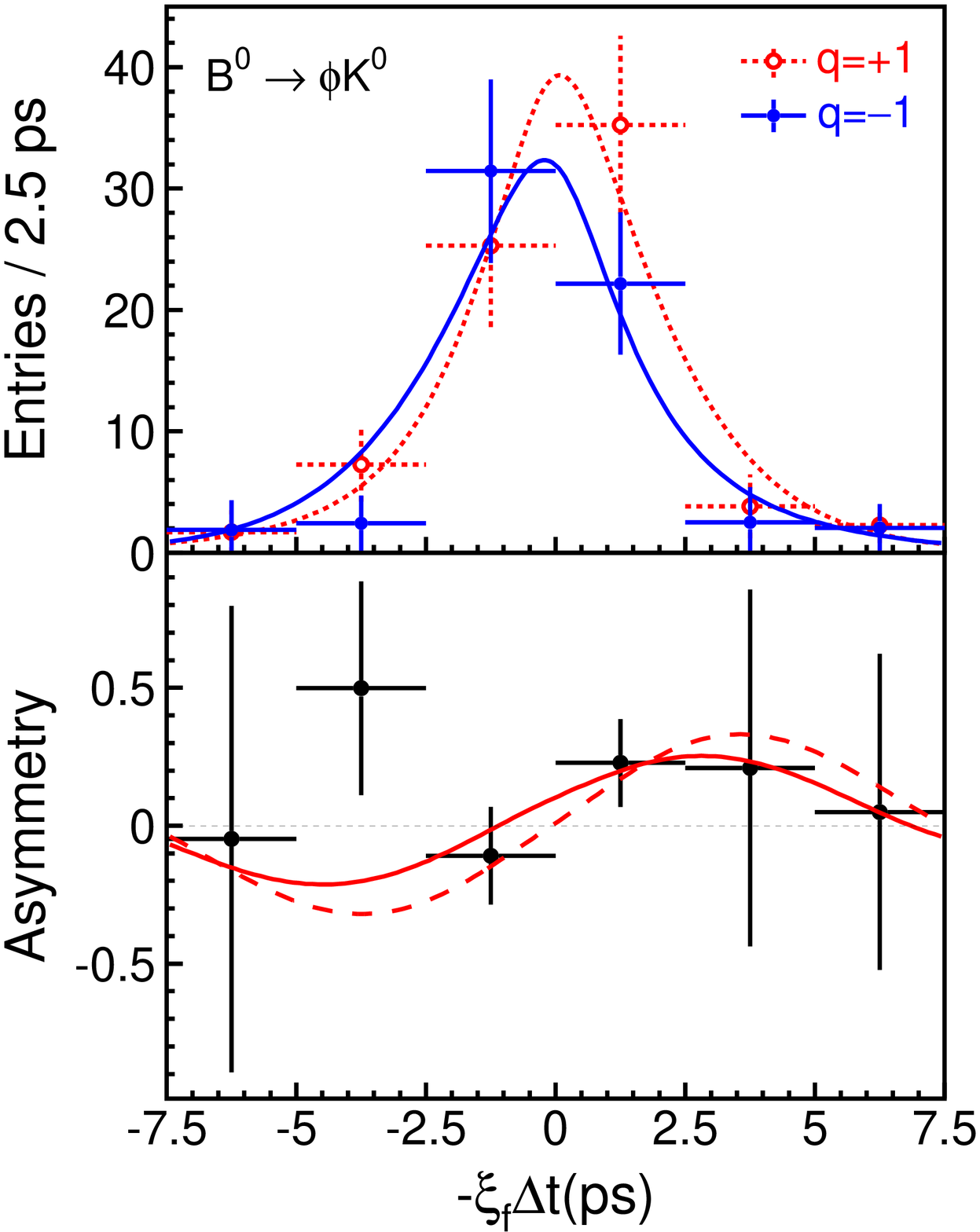}\includegraphics[width=6.0cm]{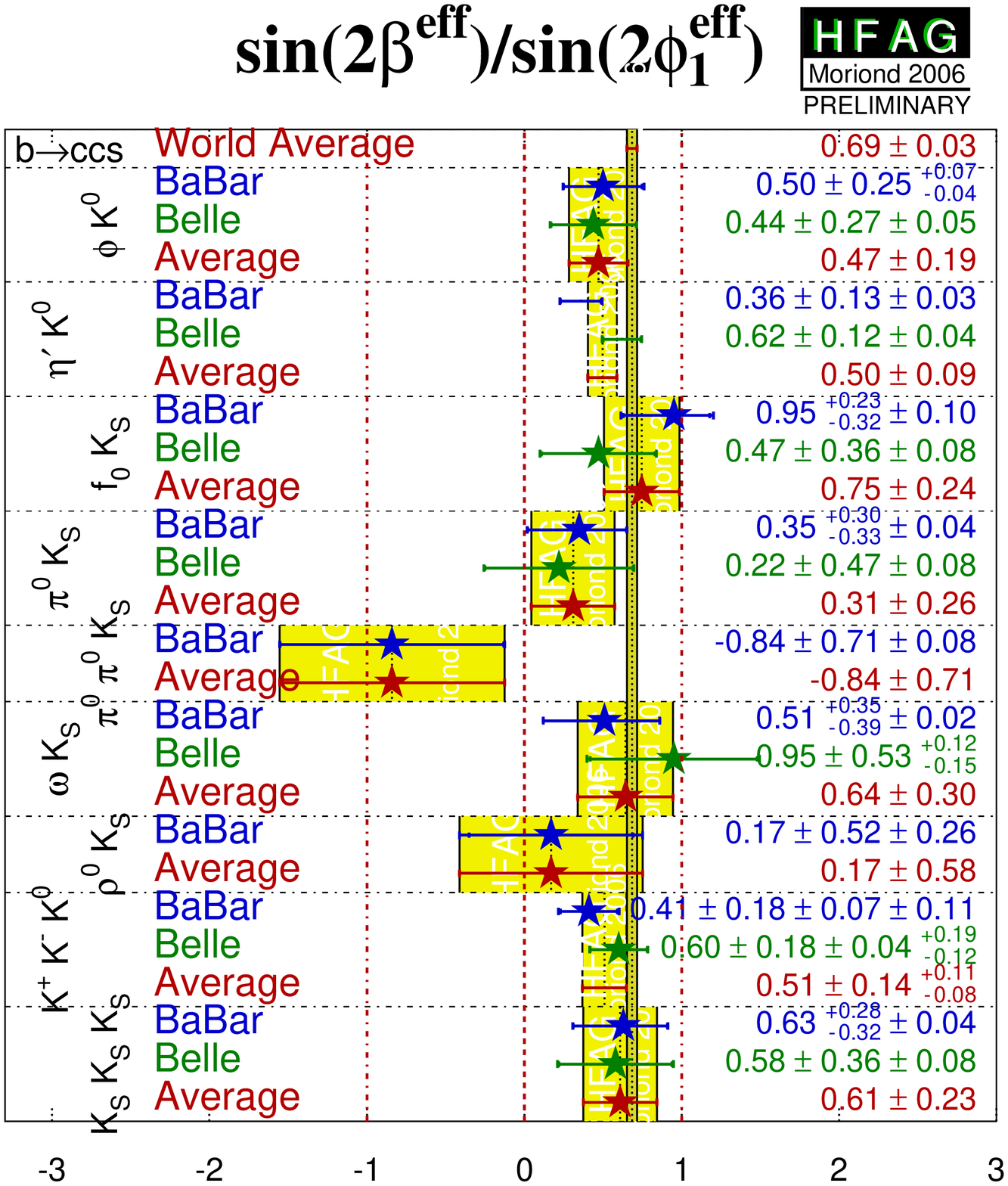}
  \caption{Left:Measurement of $S_{\Phi K^0}$ from BELLE(\cite{BELLE-beta:2}: Upper part: proper time distribution $\Delta t$ and lower part: raw asymmetry. Right: Summary of measurements of CP-violating parameters S in the $b \rightarrow q \bar q s$ modes \cite{HFAG:10}. No significant deviations from $S_{b \rightarrow c \bar c s} = sin2\phi_1$ is observed }
 \label{fig:2}
\end{figure}

In the SM final states from $b \rightarrow s \bar s s$ or $b \rightarrow d \bar d s$ transitions offer an independent test by comparing the CP-violating parameters in loop processes with those from tree-dominated ones. These decays are dominated by gluonic penguin amplitudes, but "new" non-SM physics could contribute to loop amplitudes and affect time-dependent asymmetries.
Therefore $S_{b \rightarrow s}= (sin2\phi_1^{eff})$ may deviate from its "nominal" value. One prominent pure penguin mode is $B^0 \rightarrow \Phi K^0$ which is shown in fig.~\ref{fig:2}. BELLE has combined all $b \rightarrow q \bar q s$ decay modes for 386M $B \bar B$ events and obtained
$sin2\phi_1^{eff} = 0.652\pm0.039(stat)\pm0.020(syst)$\\
With the current statistics fig.~\ref{fig:2}(right) a compilation from HFAG \cite{HFAG:10} shows no significant deviation from the "nominal" value and therefore more statistics is needed to settle this question.

\section{3. $\phi_2 (\alpha)$ measurement}

\subsection{3.1 $B \rightarrow \pi^+\pi^-$}

\begin{figure}
    \includegraphics[width=6.0cm]{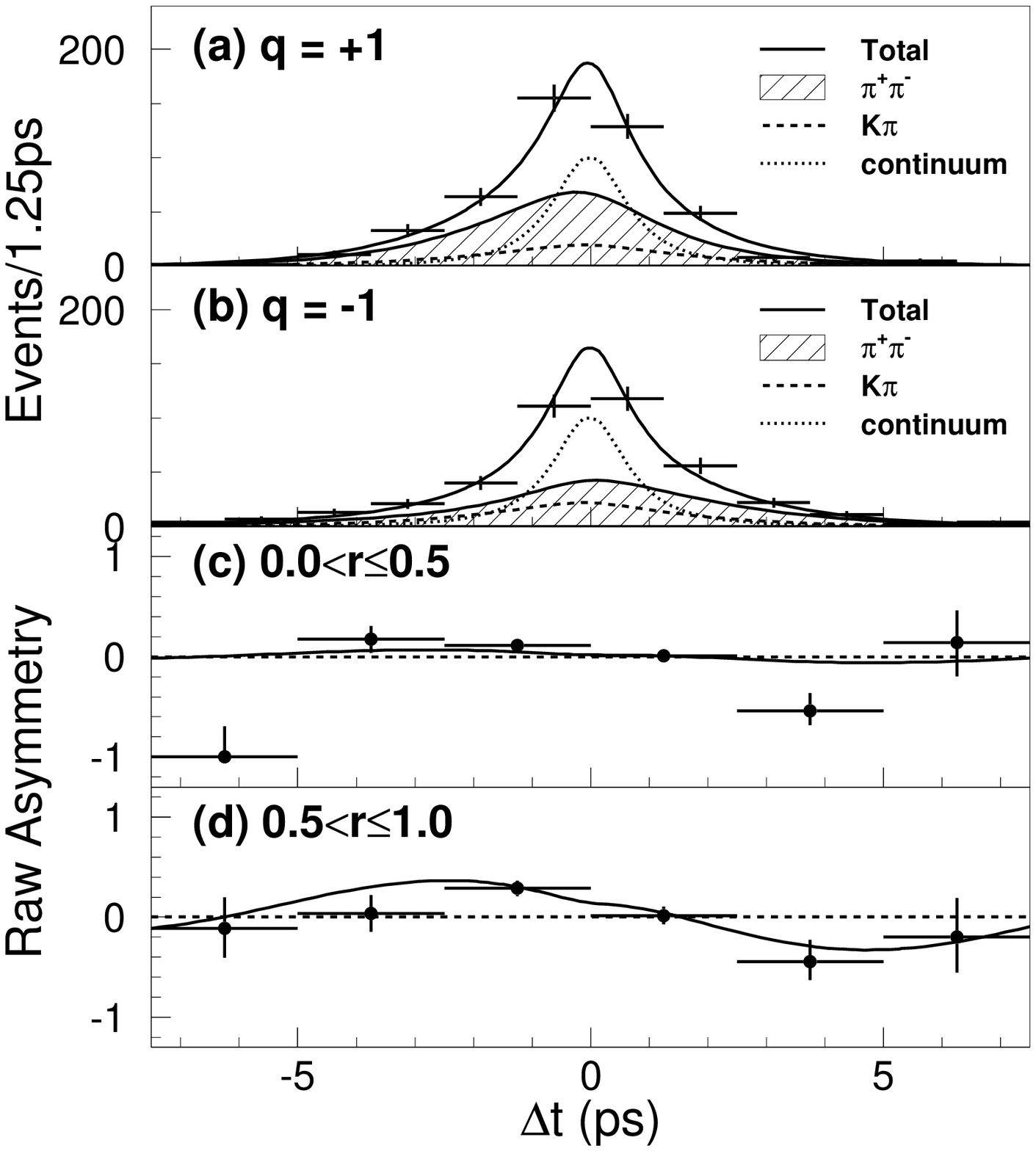}\hspace{1.0cm}\includegraphics[width=6.0cm]{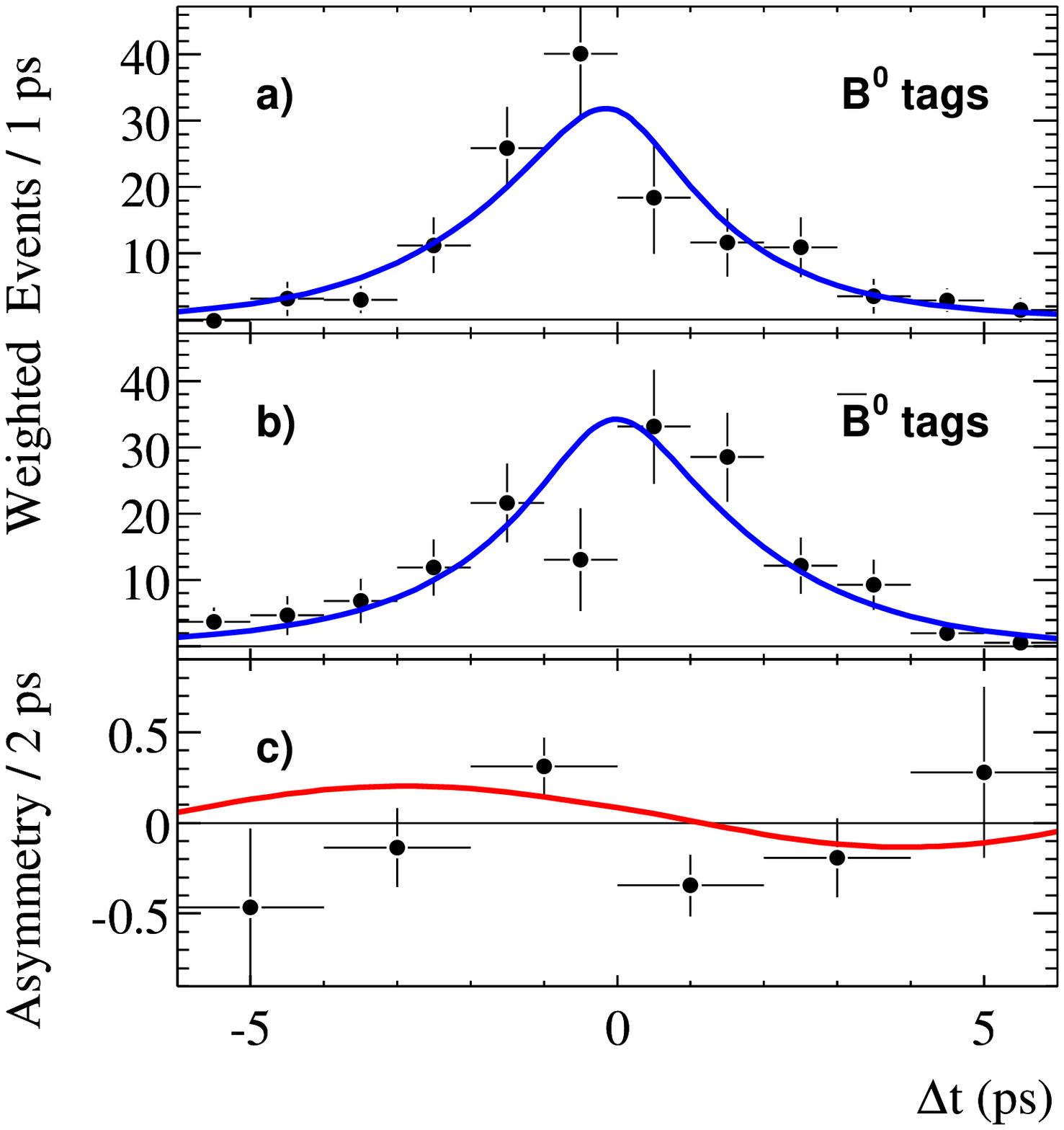}
  \caption{Measurement of $S_{\pi^+ \pi^-}$ : Upper part: proper time distribution $\Delta t$ for q=+1 ($B^0$-tag) and q=-1 ($\bar B^0$-tag) for BELLE (left) and BABAR (right). The lower part shows the raw asymmetries for both experiments, (for BELLE divided according to the purity $r$ of the tagged sample)}
 \label{fig:3}
\end{figure}

Similar as for $\phi_1$, the angle $\phi_2$ is determined by measuring the time dependent CP-asymmetries in the $b\rightarrow u $ transition. The time dependent CP-analysis has been done for $B^0 \rightarrow \pi^+\pi^-$ in both experiments. BELLE \cite{BELLE-alpha:6} gets for 666$\pm$43 signal candidates from 275M $B\bar B$ events:\\
$S_{\pi^+ \pi^-} = -0.67\pm0.16(stat)\pm0.06(syst) $   and   
$A_{\pi^+ \pi^-} =  0.56\pm0.12(stat)\pm0.06(syst) $\\
while BABAR \cite{BABAR-alpha:7} obtained from 227M $B\bar B$ events 467$\pm$33 signal candidates: \\
$S_{\pi^+ \pi^-} = -0.30\pm0.17(stat)\pm0.03(syst) $   and 
$A_{\pi^+ \pi^-} = +0.09\pm0.15(stat)\pm0.04(syst) $\\ 
as can be seen in fig.~\ref{fig:3}.
There is still a discrepancy not resolved between the two measurements; especially BELLE claims a direct CP-violation (A $\ne$ 0) with more than 4 $\sigma$ significance, while BABAR sees none. The appearance of direct CP-violation is a hint, that the contribution from $b \rightarrow d$ penguin diagrams cannot be ignored, and therefore $sin2\phi_2$ as measured in $S_{\pi^+ \pi^-}$ is different from its ideal value. 
\subsection{3.2 $B \rightarrow \rho^+\rho^-$}

\begin{figure}
    \includegraphics[width=6.0cm]{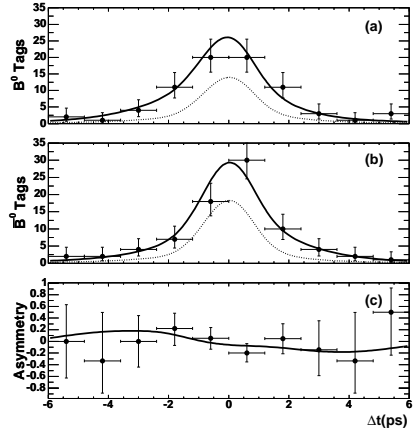}
  \caption{BABAR: Measurement of $S_{\rho^+ \rho^-}$ : Upper part: proper time distribution $\Delta t$ for $B^0$-tag and  $\bar B^0$-tag. The lower part shows the raw asymmetry}
 \label{fig:4}
\end{figure}

The decay $B^0 \rightarrow \rho^+\rho^-$ is also sensitive to this angle $\phi_2$, but it is more complicated as the final state is not a CP-eigenstate: it decays via $P \rightarrow VV$. Fortunately, the longitudinal polarization is found nearly 100 \%, as found by BELLE \cite{BELLE-rho:8} \\$f_L =0.941^{+0.034}_{-0.040}(stat)\pm0.030(syst)$ \\and therefore this final state should be a CP-eigenstate since longitudinal polarization dominates. BELLE obtains from 275M $B\bar B$ events the following values for \\
$S_{\rho^+ \rho^-} = 0.08\pm0.41(stat)\pm0.09(syst)$  and
$A_{\rho^+ \rho^-} = 0.00\pm0.30(stat)\pm0.09(syst)$ \\
while BABAR \cite{BABAR-rho:9} has measured the following values (see fig. ~\ref{fig:4}): $f_L =0.978\pm0.014(stat)^{+0.021}_{-0.029}(syst)$\\
$S_{\rho^+ \rho^-} = -0.33\pm0.24(stat)^{+0.08}_{-0.14}(syst)$  and 
$A_{\rho^+ \rho^-} = 0.03\pm0.18(stat)\pm0.09(syst)$\\
using 232M $B\bar B$ events.
Both experiments use these results for a measurement of $\phi_2 (\alpha)$, i.e.\\BELLE : $\phi_2 = (88\pm17)[deg]$  and BABAR : $\alpha = (100\pm13)[deg]$. 
The combined value from both $B \rightarrow \pi^+\pi^-$ and $B \rightarrow \rho^+\rho^-$ measurements is given in table ~\ref{tab:a}, as compiled by \cite{HFAG:10} and \cite{Fitter:14}.

\section{4. $\phi_3 (\gamma)$ measurement}

The method with the best results obtained so far uses a Dalitz plot analysis of $K_S \pi^+ \pi^-$ decay of the neutral D from the $B^\pm \rightarrow D K^\pm$ process. With the assumption of no CP-asymmetry between neutral D mesons, we can describe the amplitudes $M_+|(M_-)$ as a function of Dalitz plot variables: \\
$B^+ \rightarrow [K_S\pi^+\pi^-]K^+$    :  $M_+ =f(m^2_+,m^2_-)+re^{i(+\phi_3+\delta)}f(m^2_-,m^2_+)$\\
$B^- \rightarrow [K_S\pi^+\pi^-]K^-$    :  $M_- =f(m^2_-,m^2_+)+re^{i(-\phi_3+\delta)}f(m^2_+,m^2_-)$\\ 
where $m_+|(m_-)$ is the invariant mass of $K_S$ with $\pi^+|(\pi^-)$, $\delta$ is the strong phase, while $\phi_3$ is the weak phase to be fitted. r is the ratio of the color suppressed and allowed decays, and the amplitudes $f(m^2_+,m^2_-)$ and $f(m^2_-,m^2_+)$ were obtained from an independent $D^0$ sample of $D^{*+} \rightarrow D^0\pi^+$. Using 386M events we got $331\pm17$ $B^+ \rightarrow DK^+$,$81\pm8$ $B^+ \rightarrow D^*K^+$ and $54\pm8$ $B^+ \rightarrow DK^{*+}$ candidates. Combining all modes BELLE \cite{BELLE-gamma:11} obtained \\
$\phi_3 = (53^{+15}_{-18}(stat)\pm3(syst)\pm9(model))[deg]$ \\
where the model error reflects the uncertainty in the D-meson decay $f(m^2_-,m^2_+)$.







\section{5. Other constraints on the Unitarity triangle}

It is not only important to look at the angles of the Unitarity triangle, but there is also much interest in determining its side lengths. The two most difficult are $|V_{td}|$ and $|V_{ub}|$, i.e. transitions from third to first generation, as these are obviously very rare decays with a small branching ratio.

\subsection{5.1 $|V_{td}/V_{ts}|$ measured with the FCNC process $b\rightarrow d \gamma$}

\begin{figure}
    \includegraphics[width=12.0cm]{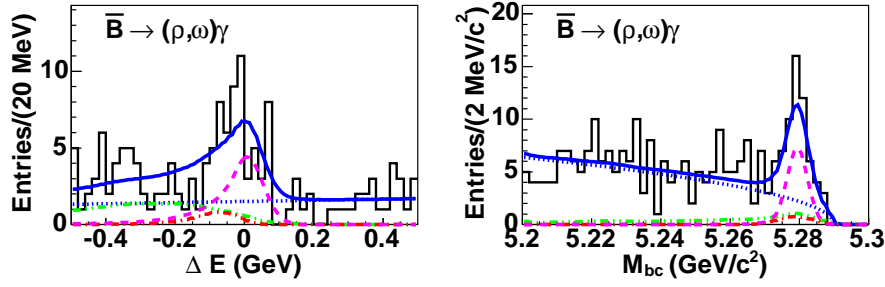}
  \caption{BELLE: Fit Results in the two projections $\Delta E$ and $M_{bc}$ for $\bar B \rightarrow (\rho,\omega) \gamma$. Curves show signal (dashed), continuum (dotted), $\bar B \rightarrow \bar K^* \gamma$ (dot-dashed), background (dot-dot-dashed) and the total fit result (solid)}
 \label{fig:6}
\end{figure}

Starting with 386M $B \bar B$ events, BELLE \cite{BELLE-dsg:12} has measured a branching fraction of Br($\bar B \rightarrow (\rho,\omega)\gamma$) = $(1.32^{+0.34}_{-0.31}(stat)^{+0.10}_{-0.09}(syst))x10^{-6}$ with 5.1 $\sigma$ significance (systematic included) with the fit of fig  ~\ref{fig:6}\\  The ratio  Br($\bar B \rightarrow (\rho,\omega)\gamma$)/Br($\bar B \rightarrow K^* \gamma$) is proportional to $|V_{td}/V_{ts}|^2$ with a kinematical factor and an SU(3) correction, and we deduce:\\
$|V_{td}/V_{ts}|$ = $0.199^{+0.026}_{-0.025}(exp)^{+0.018}_{-0.015}(theo)$.\\
Recently this quantity has also been measured by CDF \cite{CDF:15} by directly determining the oscillation amplitude $\Delta m_s$ from $B_s \bar B_s$-mixing, and they get the improved result:
$|V_{td}/V_{ts}|$ = $0.208^{+0.001}_{-0.002}(exp)^{+0.008}_{-0.006}(theo)$, consistent with BELLE.

\subsection{5.2 $|V_{ub}|$ from the purely leptonic decay $B^- \rightarrow \tau \bar\nu_{\tau}$}

\begin{figure}
    \includegraphics[width=5.0cm]{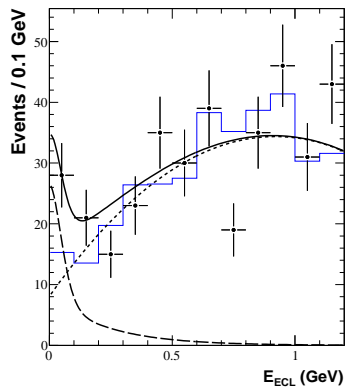}
  \caption{BELLE: Distribution of the energy in the Electromagnetic Calorimeter ($E_{ECL}$) after all selection cuts: The solid curve shows the result of the fit with the sum of the signal shape (dashed) and the background shape (dotted)}
 \label{fig:7}
\end{figure}

$|V_{ub}|$ is traditionally measured via its semileptonic decays $b \rightarrow u \ell \nu$ but this includes the knowledge of its form factor, which is difficult to obtain experimentally. Therefore BELLE has aimed to obtain this branching ratio with the advantage of a model independent measurement.
BELLE \cite{BELLE-tau:13} started with 447M B meson pairs and used only events where one side was fully reconstructed, i.e. the tag side ($B_{tag}$), and compared properties of the remaining particles, i.e. the signal side ($B_{sig}$), if this decays into $\tau$ and neutrino, using the expectations from MC for signal and background. The $\tau$ meson is identified in five decay modes ($\mu^-\bar \nu_{\mu}\nu_{\tau}$,$e^-\bar \nu_{e}\nu_{\tau}$,$\pi^-\nu_{\tau}$,$\pi^-\pi^0\nu_{\tau}$ and $\pi^-\pi^+\pi^-\nu_{\tau}$) which cover approximately 81\% of all $\tau$ decays. For all modes except $\pi^-\pi^0\nu_{\tau}$, $\pi^0$ candidates are also rejected on the signal side. Fig ~\ref{fig:7} shows the energy $E_{ECL}$ of photons, which are not associated with either $B_{tag}$ or the $\pi^0$-candidate from $\tau^-\rightarrow \pi^-\pi^0\nu_{\tau}$ Signal events should peak at low $E_{ECL}$,i.e. no photon energy, while background events show higher values from $E_{ECL}$ due to additional neutral clusters. 
We find $17.2^{+5.3}_{-4.7}$ signal events from a fit to a sample of 54  events, which means 3.5 $\sigma$ significance including systematics. Our preliminary value of the branching fraction is :\\
Br( $B^- \rightarrow \tau \bar\nu_{\tau}$) = $1.79^{+0.56}_{-0.49}(stat)^{+0.39}_{-0.46}(syst)$\\
Using the equation
Br( $B^- \rightarrow \tau \bar\nu_{\tau}$) = $\frac{G^2_Fm_Bm^2_{\tau}}{8\pi}$$(1-\frac{m^2_{\tau}}{m^2_B})^2$$f^2_B|V_{ub}|^2\tau_B$ \\
we get $f_B|V_{ub}|$ = $(10.1^{+1.6}_{-1.4}(stat)^{+1.1}_{-1.3}(syst))x10^{-4}$ GeV. 
Using the most recent HFAG \cite{HFAG:10} value from semileptonic decays for $V_{ub}=(4.39\pm0.33))x10^{-3}$ we can measure the structure constant $f_B= 229^{+36}_{-31}(stat)^{+30}_{-34}(syst)$ MeV
This is in full agreement with a recent unquenched lattice calculation with
$f_B= 216\pm 22$ MeV and thus the HFAG value for $V_{ub}$ is confirmed by this measurement.

\section{6.Summary}
Both experiments BELLE and BABAR have shown that the CKM-ansatz works extremely well for tree processes, and that this fact can be used as an "anchor point" for New Physics. Even contributions from penguin processes show no statistically significant deviation up to now, and this can be seen both in angles and side lengts of the unitarity triangle. To pin down effects from NP, statistics of the relevant processes should be increased considerably.

\begin{table}
\begin{tabular}{lrr}
\hline
  & \tablehead{1}{r}{b}{BELLE [deg.]}
  & \tablehead{1}{r}{b}{World Average [deg.]}\\
\hline
$\phi_1$ & $20.3\pm1.8$ & $21.7^{+1.3}_{-1.2}$\\
$\phi_2$ & $93^{+12}_{-11}$  & $99^{+12}_{-9}$\\
$\phi_3$ & $53^{+15}_{-18}$  & $62^{+35}_{-25}$\\
\hline
\end{tabular}
\caption{Constraints on the angles of
  Unitary Triangle}
\label{tab:a}
\end{table}

\begin{thebibliography}{9}

\bibitem{CKM:1}
M.~Kobayashi and T.~Maskawa, \emph{Prog. Theor. Phys.} \textbf{49}, 652(1973)

\bibitem{BELLE-beta:2}
K.~Abe et al. [BELLE Collaboration] \emph{Hep-ex}/0507037

\bibitem{BABAR-beta:3}
B.~Aubert et al. [BABAR Collaboration] \emph{Phys.Rev.Lett.}  \textbf{94}, 161803(2005)

\bibitem{BELLE-cos:4}
P.~Krokovny et al. [BELLE Collaboration] \emph{Hep-ex}/0605023 and \emph{Phys.Rev.Lett.}  \textbf{97}, 081801(2006)

\bibitem{BABAR-cos:5}
B.~Aubert et al. [BABAR Collaboration] \emph{Phys.Rev.D}  \textbf{71}, 032005(2005)

\bibitem{HFAG:10}
E.~Barberio et al. [HFAG: Heavy Flavor Averaging Group]
\url{http://www.slac.stanford.edu/xorg/hfag/}.

\bibitem{BELLE-alpha:6}
H.~Ishino et al. [BELLE Collaboration] \emph{Phys.Rev.Lett.}  \textbf{95}, 101801(2005)\\
For a more recent result see:\\
K.~Abe et al. [BELLE Collaboration] \emph{Hep-ex}/0608035

\bibitem{BABAR-alpha:7}
B.~Aubert et al. [BABAR Collaboration] \emph{Phys.Rev.Lett.}  \textbf{95}, 151803(2005)

\bibitem{BELLE-rho:8}
A.~Somov, A.~J.~Schwartz et al. [BELLE Collaboration] \emph{Phys.Rev.Lett.}  \textbf{96}, 171801(2006)

\bibitem{BABAR-rho:9}
B.~Aubert et al. [BABAR Collaboration] \emph{Phys.Rev.Lett.}  \textbf{95}, 041805(2005)

\bibitem{Fitter:14}
G.~Charles et al. [CKMfitter Group]
\url{http://ckmfitter.in2p3.fr/}.


\bibitem{BELLE-gamma:11}
A.~Poluektov et al. [BELLE Collaboration] \emph{Phys.Rev.D}  \textbf{73}, 112009(2006)

\bibitem{BELLE-dsg:12}
D.~Mohapatra, M.~Nakao, S.~Nishida et al.[BELLE Collaboration] \emph{Hep-ex}/0506079

\bibitem{CDF:15}
A.~Abulencia et al. [CDF Collaboration] \emph{Phys.Rev.Lett.}  \textbf{97}, 062003(2006)

\bibitem{BELLE-tau:13}
K.~Ikado et al. [BELLE Collaboration] \emph{Hep-ex}/0604018 (revised)

\end{thebibliography}
\end{document}